\documentclass[twocolumn,secnumarabic,amssymb, nobibnotes, aps, prd]{revtex4-1}

\usepackage{amsmath}
\usepackage{epsfig}
\usepackage{epic}

\usepackage{eso-pic}
\begin{document}

%\AddToShipoutPicture{\BackgroundPic} %Add background

\title{\huge  Markov chain approach to the distribution of ancestors in species of biparental reproduction} 

\author{M. Caruso}%
%\email[REVTeX Support: ]{mcaruso@ugr.es}
\affiliation{Departamento de F\'isica Te\'orica y del Cosmos, Universidad de Granada, Campus de Fuentenueva, Granada (18071), Espa\~na}

\author{C. Jarne}%
%\email[REVTeX Support: ]{jarne@fisica.unlp.edu.ar}
\affiliation{Departamento de F\'isica, Facultad de Ciencias Exactas, IFLP-CONICET, Universidad Nacional de La Plata, La Plata (1900), C.C.67, Argentina}

%\date{September 2, 2014}%

\begin{abstract}{We studied how to obtain a distribution for the number of ancestors in species of sexual  reproduction. Present models concentrate on the estimation of distributions repetitions of ancestors in genealogical trees. It has been shown that is not possible to reconstruct the genealogical history of each species along all its generations by means of a geometric progression. This analysis demonstrates that it is possible to rebuild the tree of progenitors by modeling the problem with a Markov chain. For each generation, the maximum number of possible ancestors is different. This brings huge problems for the resolution. We found a solution through a dilation of the sample space, although	 the distribution defined there takes smaller values respect to the initial problem. In order to correct the distribution for each generation, we introduced the invariance under gauge (local) group of dilations. These ideas can used to study the interaction of several processes and provide a new approach on the problem of the common ancestor. 
In the same direction, this model also provide some elements that can be used to improve models of animal reproduction.} 
\end{abstract}

\maketitle

%\tableofcontents

\section*{Introduction}

Up till now, previous attempts aiming to calculate the number of ancestors in species of sexual reproduction have not been totally successful. Present models concentrate on the estimation of distributions of ancestors repetitions in genealogical trees \cite{fis-rew,pre1,pre2}.
It has been shown that is not possible to reconstruct the genealogical history of each species along all its generations by means of a geometric progression \cite{maltusian}. The reason for that is the geometric progression is determined by a sequence of independent events. We postulate that \textit{blood relationship} is a kind of interaction that connects the events.
It is possible to re-build the tree of progenitors by modelling the problem with a Markov chain. 
If we consider a random variable which represents the number of ancestors present in a given generation, the size of the sample space depends of each generation. This brings serious complications on the solution of the problem, not only of mathematical nature. We propose submerge the original sample space into a larger one. This \textit{dilution} modifies the probability distribution. We show the need to implement a covariant derivative, due to a gauge transformation, which leaves the evolution equation invariant and correct the probability distribution. 

The main goal of present work is describe the distribution of ancestors for species with sexual reproduction, but also show the novel method used here to solve other stochastic problems.

There are two important assumptions about the biology of the considered system. The first one is that the species described here has not specific behavior of sexual partner selection (random mating reproduction) \cite{Bennet1,Bennet2}. Many species or population groups exhibit this kind of reproduction. This is the most simple case to perform the calculation. The second assumption is about the population size. The  distribution of ancestors for a given generation is contained in a population large enough to not force the selection of sexual partners blood related. The partners could be blood related or not, randomly. Current model presents a random mating in non-overlapping generations with negligible mutation and selection. These two assumptions are common to develop population genetic models, in particular, these are present in the Hardy-Weinberg principle \cite{Weinberg,Hardy}.

In this work we show a way to calculate a probability distribution to get a certain number of ancestors for each generation. We have obtained its first two cumulants: the expected value of the number and its dispersion.

Specific conditions about small size populations, or specific sexual behavior can be considered later as modifications of the general case described here.

Present work will be useful in order to understand the origin of species extrapolating the individual genealogy for all members at the beginning of the species. It is possible to go one step further, to establish how populations can be affected by certain conditions, such as isolation or migration of individuals, by studying population groups with different genetic pool \cite{Begon}. These ideas can be used to perform more realistic models in animal populations and also, improve estimations about extinction processes.

\section*{A markovian approach to the ancestors problem}

To calculate the number of ancestors of an individual it is necessary to use a statistical approach. If we simply accept that $2^{t+1}$ allow us to calculate the number of ancestors in the $t-$generation, where $t=0$ is the generation of progenitors of the first order (or parents for short) and so forth, we arrive at an absurdity. As we turn to past generations, the probability that some ancestors have been relatives is significantly larger \cite{fis-rew,pre1,pre2}. This implies a restriction on the number of ancestors with respect to $2^{t+1}$. This last quantity corresponds to the maximum possible number of ancestors in each $t-$generation.

\begin{figure}[!h]\includegraphics[width=6cm]{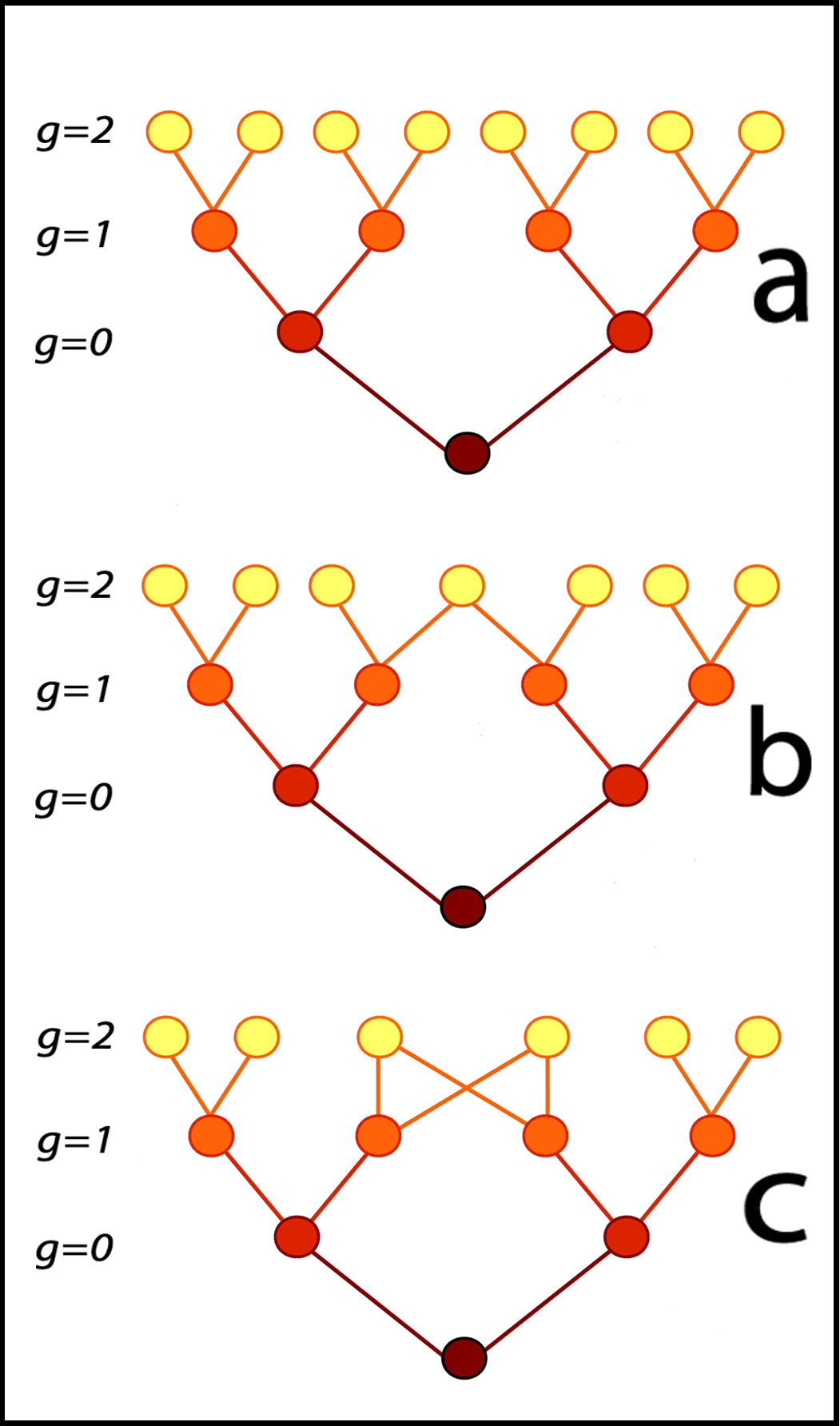}\caption{(Color online) Examples of three kinds of genealogical trees (only the first few $g-$generations) \textbf{a}: No restrictions by blood relationship. \textbf{b} and \textbf{c}: Two kind of restriction in third generation, ancestors sharing one \textbf{(b)} or two parents \textbf{(c)}. The restriction by blood relationship increases according to the degree of endogamy.}\label{f1}\end{figure}

There are several examples showing different ways in which the number of ancestors is reduced with respect to the maximum number in each generation. As stated in \cite{maltusian}, the reduction of the number of ancestors, compared to $ 2^{t+1}$, is caused by blood relationship. Figure \ref{f1} shows, as an example, only the three first generations of genealogical tree with two different ways to constrain the number of ancestors. There is a way to weight the blood relationship using a statistical approach that includes all possible kinds of relationship in each generation. In this approach the only constraint in the number of ancestors is caused by random blood relationship between individuals of the same generation. We considered a population of ancestors whose maximum size in each $t-$generation is given by the geometric progression $2^{t+1}$. 

We did not consider any restriction for the number of ancestors generated by issues related to culture, in the human case, ethological in the animal case, or isolation of populations, etc. If we want to study the distributions of ancestors of individuals from populations where there are less or equal individuals than $2^{t+1}$ for the $t-$generation, there is an additional restriction on the number of ancestors.
Blood relationship interconnects the events in the original process that leads to  $2^{t+1}$, which was generated by independent events and no relation between ancestors of each generation. 

Derrida's model \cite{fis-rew,pre1} is based on numerical simulation under the same assumptions (closed population evolving under sexual reproduction with non-overlapping generations). The population size is fixed for all generations and equally divided into two groups, representing males and females. At every generation, they form random heterosexual pairs and assign them a certain number of descendants according to a Poisson distribution. This is done by choosing for each male or female one pair of parents at random in the previous generation.

In our work the population size is not fixed, but always bigger than $2^{t+1}$ for each $t-$generation. 

We defined two random variables $y(t)$ and $x(t)$ which represents the number of individuals who are inside and outside to the set of ancestors, with respect to the maximum possible number of ancestors in each $t-$generation. For this definition we have
\begin{equation}\label{PRIMERA}
x(t)+y(t)=2^{t+1}.
\end{equation} 

We considered each generation as a link of the chain which form  a first order  Markov process \cite{Barucha,Masaaki}. This process is constructed on a given set of individuals ordered by generations. We take the current generation and we count its parents. Then we take all selected individuals and remake the previous question and so forth. There exists a generation  in which  the question or previous classification makes no more sense, in which case the process ends after a given generation. This kind of process is widely used to describe the evolution of traits that adopt a finite number of states \cite{Pagel}.

From equation \eqref{PRIMERA} we have
\begin{equation}\label{varmark}
y(t)=2^{t+1}-x(t),
\end{equation} 
if $y(t)$ describes a Markov process that implies $ x(t)$ describes another Markov process.  

We do not distinguish the different kinds of blood relationship between the ancestors of a particular generation such as brothers or cousins, and so forth. We simply consider them as indistinguishable and we just count how many there are.
For the purpose of the calculations we consider $t$ as a continuous variable. Finally we associate a discrete-time Markov process to the continuous-time Markov process $\{x(t):t\geq 0\}$ called a \textit{skeleton process} \cite{Chung} defined as $\{x(g): g \geq 0\}$, where $g$ is the generation number.

The time evolution of this process is determined by the knowledge of the probability distribution in each $t-$generation, denoted by 
\begin{equation}
p_n(t)=\mathbb{P}[x(t)=n]
\end{equation} 
for all $(n,t)\in \mathbb{S}_t\times\mathbb{R}$, where $\mathbb{S}_t$ is the sample space of $ x(t)$ which corresponds to the interval $[0,\mathfrak{n}_t]$, $\mathfrak{n}_t=\lfloor 2^{t+1}\rfloor-2$ and $\lfloor z \rfloor$ is the integer part of a real number $z$.

An equivalent way to describe the process  it is through an initial value $p_n(0)$ and the  conditional probability given by  $\mathcal{P}_{nm}(t,s)=\mathbb{P}[ x(t)=n| x(s)=m]$, which represents the transition matrix elements of the states $(m,s)\longmapsto (n,t)$.

For each generation the events are mutually exclusive. Consequently at the time  $t+\epsilon$ the probability of find $n$ restrictions is given by to the transition from $m$ restrictions at the time $t$, in this way
\begin{equation}\label{super}
p_n(t+\epsilon)=\sum_{m\in \mathbb{S}_t}\mathcal{P}_{nm}(t+\epsilon,t)\:p_m(t).
\end{equation}
After some elementary operations (see appendix section \textbf{A1}) we get 
\begin{equation}\label{evo}
d_tp_n(t)=\sum_{m\in \mathbb{S}_t}\mathtt{Q}_{nm}(t) \:p_m(t)
\end{equation}
where $d_t$ denotes the total time derivative $\frac{d}{dt}$, $\mathtt{Q}_{nm}(t)=\partial_t \mathcal{P}_{nm}(t,s)|_{s=t}$ is called the \textit{infinitesimal generator} and  $\delta_{nm}$ is the Kroneker delta. 

We define $\pmb\varphi(t)$ as an $|\mathbb{S}_t|$-tuple of the probability distribution as
$\pmb\varphi(t)=(\:p_0(t),p_1(t),\cdots,p_{\mathfrak{n}_t}(t)\:)^\intercal$, where $|\mathbb{S}_t|$ denotes the cardinal number of $\mathbb{S}_t$ and $^\intercal$ represents the transposition.

The evolution equation for the process can be expressed in a \textit{matrix form} as 
\begin{equation}\label{evolu}
d_t\pmb\varphi(t)=\pmb{\mathtt{Q}}(t)\:\pmb\varphi(t)
\end{equation}
We denote the expectation number of ancestors by $\alpha(t)=\langle y (t)\rangle$ and from the equation \eqref{varmark}
\begin{equation}
\label{med}
\alpha(t)=2^{t+1} - \langle x(t)\rangle
\end{equation}
where $\langle x^k(t)\rangle$ is the expectation value of $x(t)$ raised to the positive integer power $k$ (or $k$-moment for short) of the distribution $p_n(t)$ and by definition is $\langle x^k(t)\rangle=\sum_{n}n^k\:p_n(t)$. The quantity $\langle x(t)\rangle$ represents a constraint caused by blood relationship, which affects the expectation number of ancestors in each generation. 

\section*{Dilution of sample space via gauge group of dilations}\label{3}

The sample space of $x(t)$ is different for each $t-$generation, %\textcolor{bordo}{, i.e. we have a sample space $\mathbb{S}_t$ correlated with $t$}
thus there is enormous difficulty to solve the equation \eqref{evolu}. We considered a \textit{dilution} of $\mathbb{S}_t$ within a larger set $\mathbb{S}\supseteq\mathbb{S}_t$, for all $t$, consisting of replacing the endpoint $\mathfrak{n}_t$ by a huge number $N$. %\textcolor{bordo}{, thus each sample space is uncorrelated in time}.
 This dilution can be viewed as a \textit{dilation} represented with the substitution rule $\mathfrak{n}_t\longmapsto N$, such that $\mathbb{S}=[0,N]$. On the other hand we know that there exist a certain $T-$generation that can be considered as the end  of the process. The existence of a limit generation, $T$, allows us to choose $N=\mathfrak{n}_T$. Consequently we can solve the problem in this \textit{dilated sample space} and then recover the lost endpoint caused by the dilation through a suitable transformation. The price to pay for it is the need of renormalization of the distribution defined on $\mathbb{S}$ to compensate the dilution effect.  The renormalization takes place by a linear transformation which modifies the norm of the distribution for each generation. This \textit{local} transformation (i.e. depends of each $t$), is structured as a \textit{gauge group}, specically the \textit{group of local dilations}. Essentially, the distribution defined on $\mathbb{S}_t$ is equivalent to the renormalized distribution which is defined on the dilated sample space $\mathbb{S}$.

In summary, we can interpret that the process on $\mathbb{S}_t$ 
is the result of a process on this larger set $\mathbb{S}$ which interacts with another process on the complement of $\mathbb{S}_t$, denoted by $\mathbb{S-S}_t$. This interaction is represented by the renormalization of the distribution defined on $\mathbb{S}$, in an effective theory context.
As long as the process on $\mathbb{S}$ becomes much simpler, the description on $\mathbb{S-S}_t$ will be more complex. This is the basis for the dilation transformation, which is discussed in the appendix sections \textbf{A5} and \textbf{A6}.

We considered a version in which the sample space $\mathbb{S}_t$ is dilated to the set of natural numbers $\mathbb{N}$, including the $0$ element. Then we have only one boundary condition for the state $n=0$. This allows us to focus on \textit{time homogeneous processes}, i.e. the infinitesimal generator is independent of $t$. Another consideration is the \textit{spatial homogeneity}, i.e. the case where the infinitesimal generator does not depend on the state of the random variable $X(t)$. 

The Markov process in this larger sample space $\mathbb{N}$ requires to consider two new random variables $\{ X, Y\} $ defined on $\mathbb{N}$ and related in a similar way to the old random variables $\{ x,y\} $ from \eqref{varmark}. The associated probability distribution is denoted by $P_n(t)=\mathbb{P}\big[ X(t)=n\big]$ and defines $\pmb\phi(t)=(P_0(t),P_1(t),\cdots)^\intercal$ which satisfies the equation
\begin{equation}\label{evolucion}
d_t\pmb\phi(t)=\mathbf{Q}\:\pmb\phi(t).
\end{equation}
Knowing the initial conditions $ \pmb\phi(0)=(1,0,\cdots)^\intercal$ and the infinitesimal generator $\mathbf{Q}$ we  can write the formal solution of \eqref{evolucion} as 
 \begin{equation}
\pmb{\phi}(t)=\mathrm{exp}(t\mathbf{Q})\:\pmb{\phi}(0). 
 \end{equation}
In order to establish the matrix $\mathbf{Q}$, we study the time evolution $t\longmapsto t+\epsilon$ for small value of $\epsilon$. Therefore, only transitions to the nearest states are allowed, because the infinitesimal time evolution only has a finite variety of transition states. For $n\neq 0$ these transitions are $n\longmapsto \{n-1,n,n+1\}$ and $n\longmapsto \{n,n+1\}$, for $n=0$. 
	
Considering this brief discussion, the dynamics described by the equation \eqref{evolucion} and the imposed conditions represents a time homogeneous \textit{birth-death process}. 
A naive way to picture the process in the context of \textit{queueing theory} \cite{Kleinrock}, is through one queue representing all ancestors waiting to be classified if they are blood related or not by one server.

In the appendix section \textbf{A4}, we show how to choose a numerical matrix $\mathbf{Q}$. Finally the evolution equations takes the form
\begin{align}\label{postal}
d_tP_n(t) &= P_{n+1}(t)-2P_n(t)+P_{n-1}(t),\nonumber\\
&\\
d_tP_0(t) &= P_1(t)-P_0(t),\nonumber
\end{align}
together with the initial condition which is %in which for $t=0$ the number of ancestors is  2, i.e. 
$P_n(0)=\delta_{n0}$, we obtain the explicit solution \cite{Kleinrock}
\begin{equation}\label{solucion}
P_n(t)=e^{-2t}[I_n(2t)+I_{n+1}(2t)]
\end{equation}
where $I_n(x)$ is the modified Bessel function \cite{Abramowitz}. A brief description to obtain the solution \eqref{solucion} is also present in \cite{Kleinrock}. There is a construction of the generatrix function $g(t,z)=\sum_{n\in\mathbb{N}}P_n(t)\:z^n$, see \eqref{genfunc}, and from \eqref{evolucion} derive an equation for $g(t,z)$.

The equation \eqref{postal} is the generic expression for all Markov processes on denumerable sample spaces and continuous time with a particular values of \textbf{Q}.

\section*{The gauged distribution of ancestors}

As we have previously argued, before using this distribution to calculate the moments, it is necessary to perform a renormalization process. The reason is that the solution given by \eqref{solucion} is normalized over $\mathbb{N}$. We perform a gauge transformation \cite{Feynman}, denoted by $\mathfrak{g}_t$, which is applied to the probability distributions as
\begin{equation}\label{gauge}
\mathfrak{g}_t:P_{n}(t)\longrightarrow \lambda(t)\, P_n(t).
\end{equation}

The transformation \eqref{gauge} leaves the evolution equation \eqref{evolucion} invariant and allows both distributions to describe a Markov process. We denote $\mathfrak{p}_n(t)=\lambda(t)\; P_n(t)$ the gauge transformed distribution of $P_n(t)$. The action of the group $\mathfrak{g}_t$ %, which corresponds to multiplication by a positive function, 
applied to the distribution $ P_n (t) $ leads to a distribution $ \mathfrak {p}_n (t) $ defined over $ \mathbb{S}_t$. This idea can be understood in the context of conditional probabilities, with which we can obtain a projection of the distribution on $ \mathbb{N} $  into $ \mathbb{S}_t $, keeping the correct normalization. 

In oder words, the transformation $\mathfrak{g}_t$ leads to a new random variable $ \mathtt{X}$, which is the \textit{gauge transformed} of $X$.

To preserve the invariance of \eqref{evolucion} under $\mathfrak{g}_t$, we introduce a covariant derivative \begin{equation}\label{cov deri}
D_t=d_t-\omega(t)
\end{equation}
where $\omega(t)=d_t \lambda(t)[\lambda(t)]^{-1}$. See the appendix for a more extensive explanation.

The expectation value of $\mathtt{X}$ raised to a positive integer power $k$ is $ \langle\mathtt{X}^k(t)\rangle =\sum_{n}n^k\; \mathfrak{p}_n(t)$. 
This allows us to  write a general relation between $\langle X^k(t)\rangle$ and $\langle\mathtt{X}^k(t)\rangle$
\begin{align}
\langle \mathtt{X}^k(t)\rangle &= \lambda(t)\langle X^k(t)\rangle.
%\langle \mathtt{X}(t)^2\rangle-\langle \mathtt{X}(t)\rangle^2 &= 2^{\mathfrak{a}t+\mathfrak{b}}\left[\langle X(t)^2\rangle-2^{\mathfrak{a}t+\mathfrak{b}}\langle X(t)\rangle^2\right]
\end{align}

Rescaling the process described by $X(t)$ and using the  solution \eqref{solucion} we calculated the first two cumulants
\begin{align}\label{variansa}
\langle \mathtt{X}(t)\rangle &= \lambda(t)\langle X(t)\rangle,  \nonumber\\ 
\vspace*{2cm}&\\
\langle\:[\, \mathtt{X}(t)-\langle \mathtt{X}(t)\rangle\,]^2\,\rangle &= \lambda(t)[2t-\langle X(t)\rangle-%\textcolor{bordo}{2^{\mathfrak{a}t+\mathfrak{b}}}
\langle X(t)\rangle^2].\nonumber 
\end{align}
where 
\begin{equation}\label{medio medio}
\langle X(t)\rangle= e^{-2 t}\left[2 t\: I_1(2 t)+\left(2 t+\tfrac{1}{2}\right)\: I_0(2 t)\right]-\tfrac{1}{2}.
\end{equation}
%\surd [2^{\mathfrak{a}t+\mathfrak{b}}\{2t+\tfrac{1}{2}-e^{-2t} \left[ 2t \: I_1(2 t)+(2 t+\tfrac{1}{2})\: I_0(2 t)\right]\}-\mathfrak{c}^2(t)]^{\frac{1}{2}}

From equation \eqref{varmark} the variance of $x$ is equal to the variance of $y$. The same argument is valid for $X$ and $Y$. 

We define the \textit{standard deviation} of $\mathtt{Y}(t)=2^{t+1}-\mathtt{X}(t)$, denoted by $\sigma(t)$, as the square root of the second equation of \eqref{variansa}, which quantifies the statistical fluctuation.

As we consider a constant function $\omega(t)$, then 
\begin{equation} \label{landa}
\lambda(t)=2^{\mathfrak{a}t+\mathfrak{b}}.
\end{equation}

We have obtained a family of functions for the expectation number of ancestors
\begin{align}\label{gauged expect}
\alpha(t)=2^{t+1}-\lambda(t)\langle X(t)\rangle
%\alpha(t)&=2^{t+1}- 2^{\mathfrak{a}t+\mathfrak{b}} \left\lbrace e^{-2 t}[2 t\: I_1(2 t)+(2 t+\tfrac{1}{2})\: I_0(2 t)]
%-\tfrac{1}{2}\right\rbrace. \label{ancestros}
%\nonumber\\
%\sigma(t)&=\sqrt{ 2^{\mathfrak{a}t+\mathfrak{b}}[2t-\langle X(t)\rangle-2^{\mathfrak{a}t+\mathfrak{b}}\langle X(t)\rangle^2]}\label{varianza}
\end{align}
parametrized by the real numbers $\mathfrak{a}$ and $\mathfrak{b}$ of \eqref{landa}.

If the expected value satisfies $\alpha(t_1)=\alpha_1$ and $\alpha(t_2)=\alpha_2$, for two generations $t_1$ and $t_2$ such that $t_1\neq 0\neq t_2$, the parameters $\mathfrak{a}$ and $\mathfrak{b}$ can be obtained by
%Then we can express this two parameters as a function of two points $\mathfrak{a}=f(t_1,t_2,\alpha_1,\alpha_2)$ and $\mathfrak{b}=g(t_1,t_2,\alpha_1,\alpha_2)$

\begin{align}\label{ayb}
\mathfrak{a}=&\frac{1}{t_2-t_1}log_2 \left[\frac{2^{t_2+1}-\alpha_2}{2^{t_1+1}-\alpha_1}\frac{\langle X(t_1)\rangle}{\langle X(t_2)\rangle}\right]\nonumber\\
%&  \nonumber\\
&\\
%&  \nonumber\\
\mathfrak{b}=&\frac{1}{t_2-t_1}\Bigg\{ t_2\;log_2 \left[\frac{2^{t_1+1}-\alpha_1}{\langle X(t_1)\rangle}\right]- t_1\;log_2 \left[\frac{2^{t_2+1}-\alpha_2}{\langle X(t_2)\rangle}\right]\Bigg\}\nonumber
\end{align}
%SI NO TE GUSTA borrá lo anterior y descomentá lo siguiente: 
%\begin{align}\mathfrak{a}=&\frac{1}{t_2-t_1}log_2 \left[\frac{2^{t_2+1}-\alpha_2}{2^{t_1+1}-\alpha_1}\frac{\langle X(t_1)\rangle}{\langle X(t_2)\rangle}\right]\nonumber\\&\\ \mathfrak{b}=&\frac{1}{t_2-t_1}\left\lbrace t_2\;log_2 \left[\frac{2^{t_1+1}-\alpha_1}{\langle X(t_1)\rangle}\right]\right.\nonumber\\ &-\left. t_1\;log_2 \left[\frac{2^{t_2+1}-\alpha_2}{\langle X(t_2)\rangle}\right]\right\rbrace\nonumber \end{align}
where naturally $\alpha_i \leq2^{t_i+1}$, for $i=1,2$, to ensure good definition of $\mathfrak{a}$ and $\mathfrak{b}$. 

The gauge transformation modulates the amplitude of $\langle X(t)\rangle$. This allows us to define the notion of \textit{horizontal} and \textit{vertical range} of  $\alpha(t)$. One important point of the curve $\alpha(t)$ is the \textit{maximum generation range}, this is a nonzero generation $T$ in which $\alpha$ becomes equal to 2. Another interesting point is the maximum of $\alpha(t)$, which determines the \textit{intensity of the
process}. Without loss of generality we can choose $t_2 = T$, in which case $\alpha(t_2) = 2$, and $\alpha(t_1) = \sup \{\alpha(t): t \in [0, T]\}$. For any pair of different points, $(t_1,\alpha_1)$ and $(t_2,\alpha_2)$, considered relevant,  we select one and only one curve of the family, parameterized by $\mathfrak{a}$ and $\mathfrak{b}$ given by \eqref{ayb}. The gauge transformation $\mathfrak{g}_t$, through the $\mathfrak{a}$ and $\mathfrak{b}$ parameters, controls both  \textit{horizontal range} and  \textit{vertical range} of the process. This $T$ maybe not be a realistic value, but fix a maximum number of generations of a particular species may have.

%For illustrative purposes, in figure \ref{f2}, we have selected some curves for  $\alpha(t)$ which exemplify the possibilities of the model for a particular values of $\mathfrak{a}$ and $\mathfrak{b}$. 
%\begin{figure}[!h]\includegraphics[width=8.6cm]{fig2.png}\caption{Examples of 4 curves of the expectation value of ancestors $\alpha(t)$, given by the equation \eqref{gauged expect}, for different values of the parameters $\{\mathfrak{a},\mathfrak{b}\}$ and geometric progression $2^{t+1}$ in dashed line, which corresponds to the maximum possible number of ancestors in each $t$-generation. These 4 values of $\{\mathfrak{a},\mathfrak{b}\}$ are obtained from \eqref{ayb} and expressed in general way as $(t_1,\alpha_1,t_2,\alpha_2)=(t_1,x\,2^{t_1+1},T,2)$, where $ 0\leq x \leq 1$. In $t_1$ the curve reaches the $x$ fraction of the total number of possible ancestors for that generation, while $t_2$ defines the maximum generation range denoted by $T$. The \{green, orange, blue, red\} lines can be obtained respectively from $(t_1,\alpha_1,t_2,\alpha_2)\in\big\{(3,0.7 \times 2^{4},10,2);(3,0.9 \times 2^{4},10,2);(3,0.4 \times 2^{4},12,2);(3,0.7 \times 2^{4},12,2)\big\}$. }\label{f2}\end{figure}

\begin{figure}[!ht]\includegraphics[width=8.6cm]{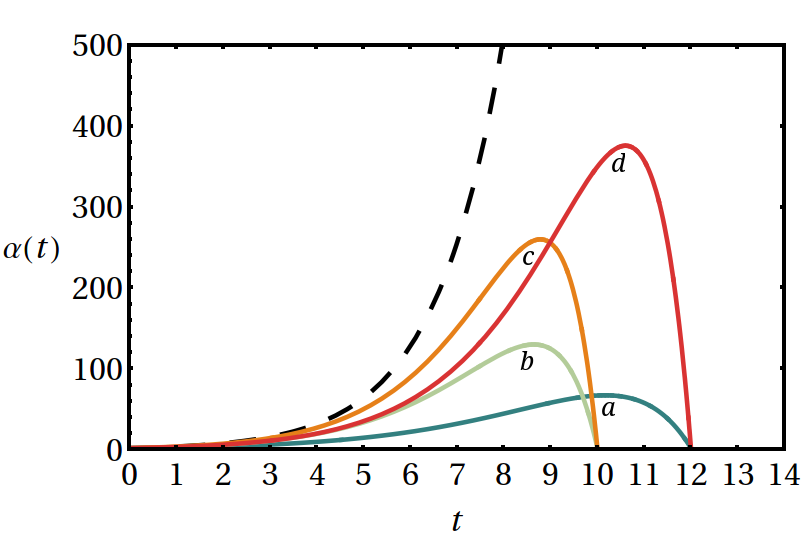}\caption{ (Color online) The 4 values of $\{\mathfrak{a},\mathfrak{b}\}$ are obtained from \eqref{ayb} and parametrized from $(t_1,\alpha_1,t_2,\alpha_2)=(t_1,\xi\,2^{t_1+1},T,2)$, where $ 0\leq \xi \leq 1$. In $t_1$ the curve reaches the fraction $\xi$ of the total number of possible ancestors for that generation, while $t_2$ defines the maximum generation range denoted by $T$. The geometric progression $2^{t+1}$ is in dash black line. The  $\{a\pmb{,}b\pmb{,}c\pmb{,}d\}$ ($\{$blue$\pmb{,}$ green$\pmb{,}$ orange$\pmb{,}$ red$\}$) lines can be obtained respectively from $(t_1,\alpha_1,t_2,\alpha_2)\in\big\{(3,0.4 \times 2^{4},12,2)\pmb{,}(3,0.7 \times 2^{4},10,2)\pmb{,}(3,0.9 \times 2^{4},10,2)\pmb{,}(3,0.7 \times 2^{4},12,2)\big\}$. }\label{f2}\end{figure}

For illustrative purposes, in figure \ref{f2}, we have selected 4 curves to the expectation number of ancestors $\alpha(t)$ given by \eqref{gauged expect} and parametrized by different values of $\{\mathfrak{a},\mathfrak{b}\}$. We include a geometric progression $2^{t+1}$, which corresponds to the maximum possible number of ancestors in each $t-$generation.

\begin{figure}[!h] 
\includegraphics[width=8.6cm]{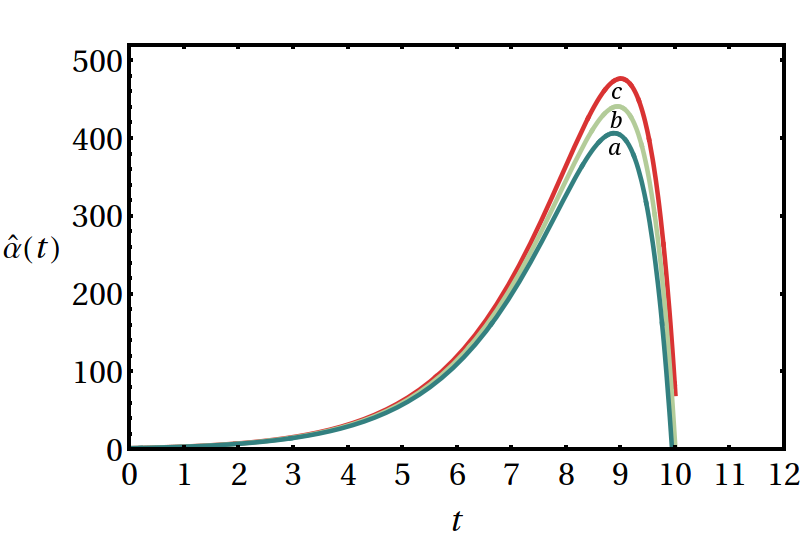}
\caption{(Color online) A band of curves, defined by the set $\mathcal{B}=\{\hat{\alpha}(t): \hat{\alpha}(t)\in[\alpha(t)-\sigma(t),\alpha(t)+\sigma(t)]\}$, contains the expectation value $\alpha(t)$ given by the equation \eqref{gauged expect} and its statistical fluctuation at 1-$\sigma$. The $\{a\pmb{,}b\pmb{,}c\}$ ($\{$blue$\pmb{,}$ green$\pmb{,}$ red$\}$) lines corresponds to the realization of $\{\alpha(t) -\sigma(t),\alpha(t),\alpha(t) +\sigma(t)\}$, with the values of $(t_1,\alpha_1,t_2,\alpha_2)=(6,0.9 \times 2^{7},10,2)$, respectively.}\label{f3}
\end{figure}

Figure \ref{f3} shows three realizations of the number of ancestors in terms of the expectation value $\alpha(t)$ and a measure of the dispersion given by $\sigma(t)$, for a particular values of $\mathfrak{a}$ and $\mathfrak{b}$.

%Finally Figure \textcolor{azul}{Citar Figura} shows the entropy define by the information theory. The entropy grows with the generation number associated to the grow of the restriction mean number. 

%\begin{figure}[!h] \centerline{\includegraphics[totalheight=5cm]{fig3.eps}}\caption{\footnotesize{A band of curves, defined by the set $\mathcal{B}=\{x(t): x(t)\in[\alpha(t)-\varepsilon(t),\alpha(t)+\varepsilon(t)]\}$, contains any possible realization of this process. \label{f3}}}\end{figure}

This model may be employed in order to recognize a possible threshold to identify high endogamic populations as well as its possible causes. Using the genealogical tree, the model can be used to indicate which living species may be near to the extinction.

\section*{Final comments and possible model extensions}

The model explained above allows to calculate the expectation number of ancestors in each generation, considering the possibility of blood relationship between individual of the same generation and 
a population of ancestors which maximum size is $2^{t+1}$. But there are two possible generalizations. The model can be extended to take into account relationships between individuals of other adjacent generations using a similar idea, simply considering higher order Markov chains. By introducing the corresponding terms in the infinitesimal generator $\mathbf{Q}$, for example an absorbing barrier \cite{Feller}, the same model can be used to calculate the expectation number of ancestors in a specific population with additional restrictions such as isolation, immigration, specific reproductive behavior, or cultural restrictions for the human case.

In future applications we could generalize the model through a new constraint to fix the maximum number of individuals at certain generation. This proposal implies a generalization of this work in which the maximum number of ancestors will be given by a piecewise function $\gamma(t)$ instead of $2^{t+1}$. This leads to slightly modify the process defined at the beginning in \eqref{varmark} as $y(t)=\gamma(t)-x(t)$ and the endpoint of the sample space for $x(t)$ be comes $\mathfrak{n}_t=\gamma(t)-2$. This generalization includes a time inhomogeneity in the infinitesimal generator $\mathbf{Q}$ and preserves an appropriate renormalization. 

We can include these possible extensions using the process $\{X(t)\}$ and study a most general gauge $\mathfrak{g}_t$ transform given by a linear transformation
\begin{equation}
\mathfrak{g}_t:\{P_m(t)\}_{m\in\mathbb{N}}\longrightarrow \mathfrak{p}_n(t)= \sum_{m\in\mathbb{N}} \lambda_{nm}(t)\; P_m(t)\label{Gauge GENERAL}
\end{equation}
where $\pmb\lambda=\{\lambda_{nm}\}$ is a non singular matrix. We can express the last expression in a matrix form
\begin{equation}
\mathfrak{g}_t:\pmb{\phi}\longrightarrow \pmb{\phi}'=\pmb{\lambda\phi}
\end{equation}
where $\pmb\phi'(t)=(\mathfrak{p}_0(t),\mathfrak{p}_1(t),\cdots)^\intercal$.

In order to preserve the invariance of \eqref{evolucion} under this generalization of $\mathfrak{g}_t$, we introduce the corresponding covariant derivative 
\begin{equation}\label{cov deri2}
D_t=d_t-\pmb{\omega}(t)
\end{equation}
where $\pmb{\omega}(t)=d_t \pmb{\lambda}(t)[\pmb{\lambda}(t)]^{-1}$. 

The evolution equation for $\pmb{\phi}'(t)$ is also invariant under the local dilation group.
\begin{align}
D_t\pmb{\phi}'(t)=\mathbf{Q}'(t)\pmb{\phi}'(t)
\end{align}
and the gauged infinitesimal generator is now  
\begin{equation}
\mathbf{Q}'(t)=\pmb{\lambda}(t)\:\mathbf{Q}\:[\pmb{\lambda}(t)]^{-1}
\end{equation}
which corresponds to a similarity transformation of $\mathbf{Q}$.

This model itself can be applied to describe other biological or physical systems with similar dynamics.
Statistical models of biparental reproduction have already been compared with physical systems before, such as, spin-glass systems \cite{0305}. In this regard the \textit{evolutionary graph theory} is an approach to study how topology affects the evolution of a population \cite{Liberman}.

Other analogous processes to the biparental reproduction in physics are described with similar statistical or markovian models \cite{Barucha}. In high energy physics the production of a cascade by a cosmic ray is described by the Heitler model \cite{heitler}. Although this model is different from the one presented here we could compare the number of ancestors with the number of particles in each generation and reinterpret this results in terms of these kinds of phenomena.

It is possible to estimate the maximum generation range, $T$, searching in the fossil record the first time that a particular species appears and use its reproductive rate. In this way we are classifying each species not in terms of life time on earth (time units), but according to the notions of generational patterns. 

The interaction of various of these processes can be combined with universal common ancestor's models \cite{Rohde} to understand the development of a certain species. The ideas in the current model can be used in biology, population ecology and genetics. An important achievement of the model is that based on the previous knowledge of the life time  of a certain species, we can calculate the number of ancestors in each generation of this species.

More fundamental uses of this ideas can be found  in mathematics, in particular in theory of stochastic processes and physics area connected with the theory of stochastic processes. Future research through a Lagrangian description may find novel applications of the present proposal. In this case we will consider the probabilities $\{p_n(t)\}$  as  the set of generalized coordinates.

\section*{Acknowledgments}

We thank our respective PhD advisors: Fernando Cornet and Mar\'ia Teresa Dova, Hern\'an Wahlberg. Also we thank to Carlos Garc\'ia Canal and Huner Fanchiotti for their helpful criticism. We are indebted to Federico Agnolin and Ben Page for reading the manuscript and providing advice. Recall also our anonymous readers and reviewers for their contribution to this work.  And finally a special mention to Micaela Moretton, Mar\'ia Clara Caruso and  Gabriel Lio for local support. 
\vspace*{0.5cm}
\section*{Appendix}
\appendix*
\vspace*{0.5cm}
%\appendix
\setcounter{equation}{0}

%%%%%%%%%%%%%%%%%%%%%%%%%%%%%%%%%%%%%%%%%%%%%%%%%%%%%%%%%%%%%%%%%%%%%%%%%%%%%%%%%%%%%%%%%%%%%%%%%%%%%%%%

\section*{\sc \textbf{A1} On the evolution equation in $\mathbb{S}$}
%solution in $\mathbb{N}$

We defined the random variable $y(t)$ associated with the number of ancestors as
	
\begin{equation}
y(t)=2^{t+1}-x(t)
\end{equation}
where $\mathbb{S}_t$ is the sample space of $x(t)$ and $t$ is a continuous variable. Then, according to a discretization process, the distribution of ancestors is obtained and the variable $t$ will be the number of generations.

We denoted by $\alpha(t)$ the expectation number of $y(t)$ as 
\begin{equation}
\label{med}
\alpha(t)=2^{t+1}-\langle x(t)\rangle
\end{equation}
where $\langle x(t)\rangle$ constrains the number of ancestors under the blood relationship hypothesis. We will focus on the process of this random variable $x(t)$ and our probability distribution denoted by $p_n(t)=\mathbb{P}[x(t)=n]$.

%The probability distribution is defined as $p_m(t):=\mathbb{P}[\mathtt{Y}(t)=m]$ and $\mathfrak{p}_n(t):=\mathbb{P}[\mathtt{X}(t)=n]$. From \eqref{varmark}, we obtain $\mathfrak{P}_m(t)=\mathfrak{p}_n(t)$, only when $m=2^{t+1}-n$. %\textcolor{amarillo}{We can say $\langle2^{t+1}\rangle=2^{t+1}$, if and only if the distribution is normalized i.e. $\sum_{n\in A} \mathfrak{p}_n(t)=\sum_{m\in B}\mathfrak{P}_m(t)=1$, where $A$ and $B$ are the sample spaces for the variables $\mathtt{X}(t)$ and $\mathtt{Y}(t)$ respectively. This will be satisfied for the final distribution. }

%The mean value of 
%\begin{align*}
%\langle\mathtt{Y}(t)\rangle %=\sum_{m\in B}m\:\mathfrak{P}_m(t)
%= \sum_{n\in S} (2^{t+1}-n)\mathfrak{p}_n(t)
%=2^{t+1}-\langle\mathtt{X}(t)\rangle
%\end{align*}

In the problem of counting ancestors the sample space $\mathbb{S}_t$ is different for each $t$-generation. For this reason we introduced a fictitious process on a larger sample space $\mathbb{S}$ than the original $\mathbb{S}_t$. Therefore we have solved the problem in $\mathbb{S}$ and renormalized the distribution to take into account of the interaction with the lost boundary of $\mathbb{S}_t$. In other words, we have considered a \textit{dilution} of $\mathbb{S}_t$ into $\mathbb{S}$. In particular we used the set of natural numbers  $\mathbb{S}=\mathbb{N}$, including the $0$ element.

We considered the study of the evolution over a generic denumerable sample space  $\mathbb{S}=[0,N]\subset \mathbb{N}$ with the random variable $X(t)$ and probability distribution $P_n(t)=\mathbb{P}[X(t)=n]$. This evolution is governed by the conditional probability given by  
\begin{equation}
\mathcal{P}_{nm}(t,s)=\mathbb{P}[X(t)=n|X(s)=m]
\end{equation}
The matrix $\mathcal{P}(t,s)$ satisfies the \textit{Chapman-Kolmogorov} equation \cite{Masaaki,Feller}
\begin{equation}
\mathcal{P}(t,s)=\mathcal{P}(t,u)\mathcal{P}(u,s)
\end{equation}
for $0\leq s \leq u \leq t$. Also the sum of the elements of each column is 
\begin{equation}\label{normalP}
\sum_{n\in \mathbb{S}}\mathcal{P}_{nm}(t,s)=1.
\end{equation}

For the general case we develop a power series of the matrix $\mathcal{P}(t+\epsilon,s)$, for a fixed value of $s$, we have
\begin{equation} \label{t evol}
\mathcal{P}_{nm}(t+\epsilon,s)=\mathcal{P}_{nm}(t,s)+\epsilon\:\partial_t\mathcal{P}_{nm}(t,s)+\hdots
\end{equation}
where $\partial_t$ is a simplified notation of partial time derivative $\frac{\partial}{\partial t}$ 

In order to obtain the equation \eqref{super} we study the time evolution $t\longmapsto t+\epsilon$, for small value of $\epsilon$. We need to know $\mathcal{P}_{nm}(t+\epsilon,t)$ then \eqref{t evol} becomes 
\begin{equation}\label{inf t evol}
\mathcal{P}_{nm}(t+\epsilon,t)=\delta_{nm}+\epsilon\:\partial_t\mathcal{P}_{nm}(t,s)|_{s=t}+\hdots%+\epsilon\:\mathrm{Q}_{nm}(t)+\hdots
\end{equation}
we recognize the second term of \eqref{inf t evol} as the infinitesimal generator $\mathrm{Q}_{nm}(t)$ \begin{equation}\label{inf t evol}
\mathrm{Q}_{nm}(t)=\lim_{\epsilon \to 0}\frac{\mathcal{P}_{nm}(t+\epsilon,t)-\delta_{nm}}{\epsilon}	
\end{equation}

%The evolution of this kind of Markov process can be studied through \begin{equation}\label{genevol}P_n(t+\epsilon)=\sum_{k}\mathcal{P}_{nk}(t+\epsilon,t)P_k(t)\end{equation} which is similar to the equation \eqref{super} of the main paper.

We assumed that $t$ is continuous. This allows us to evaluate the process at any $t$ between two generations, but it also reduces the number of possible states in an infinitesimal evolution. Therefore only transitions to the nearest states are allowed. For the ancestry problem, the infinitesimal time evolution has a finite number of transition states. These transitions are denoted by $n\longmapsto n'$, where $n'\in T_n$, i.e. $n'$ depends on the initial state $n$. We write explicitly  $T_0=\{0,1\}$, $T_N=\{N-1,N\}$ and for $n\neq 0,N$: $T_n=\{n-1,n,n+1\}$.  Furthermore, if $|n'-n|>1$ the corresponding transition probability  is zero.

From equation \eqref{normalP} the matrix $\mathcal{P}(t+\epsilon,t)$ is normalized for all $t$ and small $\epsilon$ as 
\begin{equation}\label{normalP2}
\sum_{n'\in T_n}\mathcal{P}_{n'n}(t+\epsilon,t)=1
\end{equation}
note that $n'$ runs over $T_n$, depending on whether $n$ is equal to $0$, $N$ or any other value of $\mathbb{S}-\{0,N\}$. 

We express $\mathcal{P}_{n'n}(t+\epsilon,t)$  for these three cases from \eqref{inf t evol}
\begin{align}\label{neq0}
\mathcal{P}_{n-1\,n}(t+\epsilon,t)&=\mu_n(t)\epsilon + \mathcal{O}_t(\epsilon),\nonumber\\
\mathcal{P}_{n+1\,n}(t+\epsilon,t)&=\nu_n(t)\epsilon+\mathcal{O}_t(\epsilon),\\
\mathcal{P}_{n'n}(t+\epsilon,t)&= 0, \quad |n'-n|>1,\nonumber\\
\mathcal{P}_{nn}(t+\epsilon,t)&= 1-[\nu_n(t)+\mu_n(t)]\epsilon+\mathcal{O}_t(\epsilon),\nonumber
\end{align}
where $\mathcal{O}_t(x)$ represents a type of function that goes to zero with $x$ faster than $x$, for a given $t$, that is
\begin{equation}
\lim_{x\to 0}\frac{\mathcal{O}_t(x)}{x}=0.
\end{equation}

The fourth equation of \eqref{neq0} is obtained through the first three of them. In the general case we proceed as follows from \eqref{normalP}
\begin{equation}
\mathcal{P}_{n\,n}(t,s)=1-\sum_{n'\in \mathbb{S}-\{n\}}\mathcal{P}_{n'n}(t,s)
\end{equation}
then from an infinitesimal time evolution $s\equiv t\longmapsto t+\epsilon$ and \eqref{normalP2} 
\begin{align*}
\mathcal{P}_{nn}(t+\epsilon,t)&=1-\sum_{n'\in T_n-\{n\}}\mathcal{P}_{n'n}(t+\epsilon,t)\\
\mathcal{P}_{nn}(t+\epsilon,t)&=1-\mathcal{P}_{n+1\,n}(t+\epsilon,t)-\mathcal{P}_{n-1\,n}(t+\epsilon,t)
\end{align*}
which is the fourth equation of \eqref{neq0}.

Replacing \eqref{neq0} in %\eqref{genevol} 
\eqref{super} and  written for $n\neq 0$, we have
\begin{align*}
P_n(t+\epsilon)&=[\nu_{n-1}(t)\epsilon+\mathcal{O}_t(\epsilon)]\:P_{n-1}(t)\nonumber\\
&+[\mu_{n+1}(t)\epsilon+\mathcal{O}_t(\epsilon)]\:P_{n+1}(t)\\
&+[1-\nu_n(t)\epsilon-\mu_n(t)\epsilon+\mathcal{O}_t(\epsilon)]\:P_n(t)\nonumber
\end{align*}
then 
\begin{align*}
\frac{P_n(t+\epsilon)-P_n(t)}{\epsilon}=
 &\left[\nu_{n-1}(t)+\frac{\mathcal{O}_t(\epsilon)}{\epsilon}\right]P_{n-1}(t)\nonumber\\
+&\left[\mu_{n+1}(t)+\frac{\mathcal{O}_t(\epsilon)}{\epsilon}\right]P_{n+1}(t)\\
-&\left[\nu_n(t)+\mu_n(t)+\frac{\mathcal{O}_t(\epsilon)}{\epsilon}\right]P_n(t)\nonumber
\end{align*}
taking the limit $\epsilon \to 0$
\begin{align}\label{posta}
d_tP_n(t)= &\:\nu_{n-1}(t)P_{n-1}(t)+\:\mu_{n+1}(t)P_{n+1}(t)\nonumber\\
           &\\          
          -&[\nu_n(t)+\mu_n(t)]P_n(t)\nonumber
\end{align}

The stochastic process described by equation \eqref{posta} corresponds to the general class of stochastic dynamics called \textit{birth and death process}, which includes the \textit{queueing process} \cite{Masaaki,Kleinrock}. 

The functions $\mu_n(t)$ and $\nu_n(t)$ are part of the infinitesimal generator $\mathbf{Q}(t)$. From the first equation of \eqref{neq0} we have 
\begin{equation}
\mu_n(t)=\lim_{\epsilon\to0}  \frac{\mathcal{P}_{n-1\,n}(t+\epsilon,t)}{\epsilon}
\end{equation}
which is exactly the element $\mathrm{Q}_{n-1\,n}(t)$ of \eqref{inf t evol}. 

In summary we list all the elements of $\mathbf{Q}(t)$ 
\begin{align}\label{Qelements}
&\mathrm{Q}_{nn}(t)=-[\nu_n(t)+\mu_n(t)], \quad \mathrm{Q}_{n-1\,n}(t)=\mu_n(t) \nonumber\\	
&\\
&\mathrm{Q}_{n+1\,n}(t)=\nu_n(t),\quad \mathrm{Q}_{n'n}(t)=0, \quad |n'-n|>1\nonumber
\end{align}

We write \eqref{posta} in a matrix form as
\begin{equation}\label{evoluciont} 
d_t\pmb{\phi}(t)=\mathbf{Q}(t)\pmb{\phi}(t)
\end{equation}
where $\pmb{\phi}(t)=(P_0(t),\cdots,P_N(t))^\intercal$ and $\mathbf{Q}(t)=\{\mathrm{Q}_{n'n}(t)\}$ is given by \eqref{Qelements}.

It should also be pointed out that the coefficients $\mu_0$ and $\nu_N$ must be zero, otherwise we require more states than $[0,N]$ in $\mathbb{S}$. In other words, if $\mu_0$ or $\nu_N$ are not equal to zero the left side of \eqref{normalP2} is not equal to one.

%%%%%%%%%%%%%%%%%%%%%%%%%%%%%%%%%%%%%%%%%%%%%%%%%%%%%%%%%%%%%%%%%%%%%%%%%%%%%%%%%%%%%%%%%%%
\section*{\sc \textbf{A2} Homoegeneous Hypothesis}

In this section we show how the hypothesis of spatial and temporal homogeneity are used working in the dilated sample space.

We have already said that if there exists a certain $T-$generation that can be considered as the stop of the process, the upper limit of the dilated space $\mathbb{S}$, denoted by $N$, can be chosen as 
\begin{equation}
N=\mathfrak{n}_T,
\end{equation}
where $\mathfrak{n}_T=\sup\{n:n\in\mathbb{S}_t, \forall t\in [0,T]\}$. 

Also we considered a dilution of $\mathbb{S}_t$ into $ \mathbb{N}$, i.e. the generic dilated space $\mathbb{S}$ is equal to $\mathbb{N}$, or $N\longrightarrow \infty$. This assumption is true from $\mathbb{S}_t\subseteq\mathbb{N}$, no matter how big is $\mathfrak{n}_T$. In this case the space-time on the process is infinite and we have an infinitesimal generator on $\mathbb{N}$ independent of the state of the random variable $X$ and time-independent, and the process is \textit{space-time homogeneous}.

However the effect of these hypotheses can be compensated with an interaction with the process on $\mathbb{N-S}_t$, see the section \textbf{A4} and for more details.

Essentially we will say that the renormalized distribution defined on $\mathbb{N}$ is equivalent to the distribution defined on $\mathbb{S}_t$. This equivalence is based on the invariance of evolution equation. In this way both distribution corresponds to a Markov process, see section \textbf{A4}.
%%%%%%%%%%%%%%%%%%%%%%%%%%%%%%%%%%%%%%%%%%%%%%%%%%%%%%%%%%%%%%%%%%%%%%%%%%%%%%%%%%%%%%%%%%%
\section*{\sc \textbf{A3} On the space-time homogeneous solution in $\mathbb{N}$}

The evolution equation in the dilated sample space $\mathbb{N}$ under the space-time homogeneity is
\begin{equation}\label{evolucionN}
d_t\pmb{\phi}(t)=\mathbf{Q}\pmb{\phi}(t),
\end{equation}
with $\pmb\phi(t)=(P_0(t),P_1(t),\cdots)$ and identify the matrix $\mathbf{Q}$ as 

\begin{align}\label{gen inf}
\mathbf{Q}=\left(\begin{matrix}
-\nu &   \mu     &     0    & & \hdots\\
  \hspace*{0,22cm}\nu & -\mu-\nu  &    \mu   & & \hdots\\
 \hspace*{0,22cm}  0  &   \nu     &-\mu-\nu  & & \hdots\\
 \hspace*{0,22cm}  0  &    0      &  \nu     & & \hdots\\
 \hspace*{0,22cm}  0  &    0      &     0    & & \hdots\\
 \hspace*{0,22cm}\vdots&\vdots&\vdots&       &\ddots&
\end{matrix}\right).
\end{align}

%\textcolor{bordo}{REPENSAR el concepto de tiempo de \textbf{ingreso} y \textbf{egreso} denotados por \textbf{T}$_\pm$. }
Under the initial condition  $\pmb\phi(0)=(1,0,\hdots)^\intercal$, the solution of \eqref{evolucionN} with \eqref{gen inf} is 
\begin{align}\label{soluzion}
P_n(t)&=e^{-(\nu+\mu) t}\Bigg[\rho^{n/2}I_n(\zeta\:t)+\rho^{(n-1)/2} I_{n+1}(\zeta\:t)
\nonumber\\
&\\
&+(1-\rho)\rho^n\sum_{j=n+2}^\infty \rho^{-j/2}I_j(\zeta\:t)\Bigg]\nonumber.
\end{align}
where $I_n(x)$ is the modified Bessel function \cite{Abramowitz}, $\rho=\nu/\mu$ and $\zeta=2\sqrt{\nu\mu}$. The first solution of \eqref{evolucionN} appeared in the 1950's, see \cite{Bailey,Champernowne,Clark,Lederman}. A description to obtain the solution \eqref{soluzion} is also presented in \cite{Kleinrock}.

%%%%%%%%%%%%%%%%%%%%%%%%%%%%%%%%%%%%%%%%%%%%%%%%%%%%%%%%%%%%%%%%%%%%%%%%%%%%%%%%%%%%%%%%

\section*{\sc \textbf{A4} Moments of the distribution}

In this section we calculate the first two cumulants of the distribution obtained in \eqref{soluzion}. 

First of all we demonstrated the existence of all $k$-moment of the distribution $P_n(t)$ defined by
\begin{equation}\label{mom}
\langle X^k(t)\rangle=\sum_{n\in\mathbb{N}}n^k\:P_n(t)
\end{equation}
with $k\in\mathbb{N}$. It is possible to demonstrate that all series defined above converge uniformly  $\forall t$.  To demonstrate this, we define the \textit{generatrix function}
\begin{equation}\label{genfunc}
g(t,z)=\sum_{n\in\mathbb{N}}P_n(t)\:z^n
\end{equation}
for $z\in\mathbb{R}$. If it converges, $g(t,z)$ is well defined. For our case, we know that 

\begin{equation}
\sum_{n\in\mathbb{N}} I_n(x)\longrightarrow \tfrac{1}{2}[e^x+I_0(x)],
\end{equation} 
converges uniformly \cite{Abramowitz}. This allows us to write the distribution's norm and demonstrate that converge uniformly 
\begin{equation}
\sum_{n\in\mathbb{N}} P_n(t)\longrightarrow 1.
\end{equation}
Another argument for the general birth-death process, based on the nature of the coefficients $\{(\nu_n,\mu_{n+1}): n\in \mathbb{N}\}$  leads us to the same conclusion \cite{Feller}.

For each $t$, we demonstrated that $g(t,1)\longrightarrow 1$ uniformly, then for the Abel's theorem  \cite{Apostol} 
\begin{equation}
\sum_{n\in\mathbb{N}}P_n(t)\:z^n \longrightarrow g(t,z) 
\end{equation}
uniformly for each  $z\in[0,1]$. 

Therefore the equation \eqref{genfunc} can  be derived term by term keeping the uniform convergence $\forall (t,k)$
\begin{equation}\label{series}
\sum_{n\in\mathbb{N}} \frac{n!}{(n-k)!}P_n(t)\longrightarrow\partial_{z^k}g(t,z)|_{z=1}.
\end{equation} 

Finally $\langle X^k(t)\rangle$ can be obtained as a combination of the series given by \eqref{series},  which converges to $\partial_{z^k}g(t,z)|_{z=1}$, and this completes the demonstration.

We can use two methods for the calculation of the first two moments, given the solution \eqref{soluzion}. 

The first one is by definition \eqref{mom} and using the identity \cite{Abramowitz}  
\begin{equation}
nI_n(x)=\frac{x}{2}[I_{n-1}(x)-I_{n+1}(x)].
\end{equation}

The second one is based on the series \eqref{mom} converges uniformly. We can derive term by term the series \eqref{mom}. Using the evolution equation \eqref{evolucionN} with \eqref{gen inf}, we can obtain a differential equation for the expectation value $\langle X(t) \rangle$ and $\langle X^2(t) \rangle$ 
\begin{align}
d_t\langle X(t) \rangle&=(\nu-\mu)+\mu P_0(t),\\
&\nonumber\\
d_t\langle X^2(t) \rangle&=2\nu -d_t\langle X(t) \rangle+2(\nu-\mu)\langle X(t) \rangle .
%d_t\sigma(t)&=\nu+\mu+2(\nu-\mu)\langle X(t) \rangle-d_t\langle X(t) \rangle-\langle X(t)\rangle^2.
\end{align}

For the initial condition $P_n(0)=\delta_{n0}$ then $\langle X^k(0) \rangle=0$ and integrate the last two equations
\begin{align}
\langle X(t) \rangle&=(\nu-\mu)t+\mu\int_0^t P_0(\tau)d\tau, \label{medio}\\
\langle X^2(t) \rangle&=2\nu t-\langle X(t) \rangle +2(\nu-\mu)\int_0^t \langle X(\tau) \rangle d\tau . \label{varianza}
\end{align}

The expression \eqref{medio} shows that to determine $\langle X(t) \rangle$ is sufficient to know the distribution of probability of no blood relationship $P_0(t)$ from \eqref{soluzion} . Also the expression \eqref{varianza} shows that to determine $\langle X^2(t)\rangle$ is sufficient to know $\langle X(t) \rangle$. 

This two methods arrives at the same result for $\langle X(t)\rangle$.

We are interested in computing the first two cumulants. The first one is the expectation value and the second one is defined as function of the first two moments by $\langle [X(t)-\langle X(t)\rangle ]^2\rangle=\langle X^2(t)\rangle-\langle X(t)\rangle^2$ respectively.

We have said that this is due to dilation of the sample space $\mathbb{S}_t \longmapsto \mathbb{N}$. 
This implies that distributions defined on $\mathbb{N}$ takes smaller values than they should take on $\mathbb{S}_t$.  

Also we show how $\langle X(t) \rangle $ is small compared to $2^{t+1}$, for any value of $\nu\geq 0$ and $\mu\geq 0$.

Using \eqref{medio} and $0\leq P_0(t)\leq  1$ we can express 
\begin{equation}
|\langle  X(t) \rangle|\leq |\nu-\mu|t+\mu\int_0^t P_0(\tau)d\tau\leq |\nu-\mu|t+\mu t.
\end{equation}
This shows that $\langle X(t)\rangle$ is subordinated to a linear function in $t$, for all $\mu,\nu$, then the expected value of ancestors $\langle  Y(t) \rangle=2^{t+1}-\langle  X(t) \rangle$ it will grow indefinitely with $t$ as $2^{t+1}$. We can consider by ignorance $\nu=\mu$, in the sense of not knowing the functional form of the trend of the number of ancestors. Although the knowledge of any particular trend can be introduced in the gauge transformation.  And finally without loss of generality we can take $\mu=1=\nu$, since the problem of non saturation will be solved by dilate the distribution to compensate for the dilution of the sample space, as we shall see in the next section. This dilation can be seen as a renormalization. In the first place we considered the case where $X(t)$ is defined on $\mathbb{N}$ with no renormalization at all. In this way, the renormalization is interpreted as an operation where the correct scale of the interaction is retrieved modifying $\langle X (t) \rangle$, \textit{as if} we had solved the problem in the original sample space $\mathbb{S}_t$.

%%%%%%%%%%%%%%%%%%%%%%%%%%%%%%%%%%%%%%%%%%%%%%%%%%%%%%%%%%%%%%%%%%%%%%%%%%%%%%%%%%%%%%%%%%%%%%%%%%%%%%%%

\section*{\sc \textbf{A5} On the interaction with a ficticious enviroment}

We analyzed one of the main concepts: the dilution of the sample space involves the study of an interaction between the initial sample space with the ficticious enviroment. 
The introduction of the dilution takes into account the interaction with the lost boundary of $\mathbb{S}_t$, under the condition of $\mathbb{S}$ is large enough to include $\mathbb{S}_t$ for all $t$. 

We can view the process on $\mathbb{S}_t$ as the result of a process on $\mathbb{S}$ which interact with another process defined on the complement set $\mathbb{R}_t=\mathbb{S-S}_t$. This point of view can be described in a mathematical precise sense. We define the associated vector spaces $\{S,S_t,R_t\}$ to the sets $\{\mathbb{S,S}_t,\mathbb{R}_t\}$, where $\pmb{\phi}$ is a vector in $S$, which $dim(S)=N+1$. If the sample space $\mathbb{S}_t$ has $\mathfrak{n}_t+1$ number of states, we define $\pmb\varphi$ as the first $\mathfrak{n}_t+1$ component of $\pmb{\phi}$, i.e. $\pmb{\varphi}$ is vector of $S_t$. We expressed these vectors in a canonical basis $\{\pmb{e}_n\}_{n=0,\,\cdots,\,N}$ such that 
\begin{align}
\pmb{\phi}(t)&=\sum_{n\in \mathbb{S}}\,P_n(t) \,\pmb{e}_n\\
\pmb{\phi}(t)&=\sum_{n\in \mathbb{S}_t}\,P_n(t) \,\pmb{e}_n+\sum_{n\in \mathbb{R}_t}\,P_n(t) \,\pmb{e}_n\nonumber
\end{align}
then we have
\begin{align}\label{dynS}
\pmb{\phi}(t)=\pmb{\varphi}(t)+\pmb{\psi}(t)
\end{align}
where $\pmb{e}_0=(1,0,\cdots,0)^\intercal $, $\pmb{e}_1=(0,1,\cdots,0)^\intercal $ and so forth. For construction $S_t$ is orthogonal to $R_t$. The dimensions of these vector spaces are determined by $dim(S)=N+1$ and $dim(S_t)=\mathfrak{n}_t+1$. 

From the equation \eqref{dynS} we see, roughly speaking, that the process on $\mathbb{S}_t$ is the result of interaction between  the process on $\mathbb{S}$ and $\mathbb{R}_t$ through 
\begin{align}
\pmb{\varphi}(t)=\pmb{\phi}(t)-\pmb{\psi}(t)
\end{align}

We can write down the evolution equation \eqref{evoluciont} in the form 
\begin{align}
d_t\pmb{\varphi}(t)&=\mathbf{Q}_{ss}(t)\pmb{\varphi}(t)+\mathbf{Q}_{sr}(t)\pmb{\psi}(t) \label{evolucion partida1}\\
\nonumber \\
d_t\pmb{\psi}(t)&=\mathbf{Q}_{rs}(t)\pmb{\varphi}(t)+\mathbf{Q}_{rr}(t)\pmb{\psi}(t)\label{evolucion partida2}
\end{align}
where $\mathbf{Q}_{ab}$ is the $a\times b$ \textit{block matrix} of $\mathbf{Q}$, for $a,b\in \{s,r\}$, $s=\mathfrak{n}_t+1$ and $r=N-\mathfrak{n}_t$, explicitly 
\begin{align}\label{gen inf2}
\mathbf{Q}=\left(\begin{matrix}
\mathbf{Q}_{ss} &\mathbf{Q}_{sr}\\
\mathbf{Q}_{rs} &\mathbf{Q}_{rr}
\end{matrix}\right)
\end{align}

We see that the equations for $ \pmb{\varphi} $ and $ \pmb{\psi} $ are coupled, hence the interaction character which was noted above. To be more specific, we write a relation of two \textit{partial solutions} $\pmb{\psi}$ and $\pmb{\varphi}$
\begin{align}\label{interact rs}
\pmb{\psi}(t)&=\int_\mathbb{R} \mathbf{K}_{rs}(t,t')\pmb{\varphi}(t')dt' 
\end{align}
where \eqref{interact rs} satisfies  \eqref{evolucion partida2} and  $\mathbf{K}_{rs}(t,t')$ is the kernel of this transformation defined as
\begin{equation}\label{kernelrs}
\mathbf{K}_{rs}(t,t')=\mathcal{T}\bigg\{\exp\bigg[\int_{t'}^t \mathbf{Q}_{rr}(\tau)d\tau\bigg]\bigg\}\:\mathbf{Q}_{rs}(t')\:\theta(t-t')
\end{equation}
and $\theta(t)$ are the \textit{unit step} distribution and $\mathcal{T}$ is the time-ordered operator defined as
\begin{align}
\mathcal{T}[\mathbf{A}(t)\mathbf{A}(u)]=\left\lbrace\begin{matrix}
\mathbf{A}(t)\mathbf{A}(u)&:& t>u\\
\mathbf{A}(u)\mathbf{A}(t)&:& u>t
\end{matrix}\right.
\end{align}

We can obtain another expression for equation \eqref{evolucion partida1}
\begin{align} \label{evolucion varphi}
d_t\pmb{\varphi}(t)&=\int_\mathbb{R} \mathbf{K}_{ss}(t,t')\pmb{\varphi}(t')dt'
\end{align}
where $\mathbf{K}_{ss}$ is the kernel of the integro-differential equation \eqref{evolucion varphi} defined as  
\begin{equation}\label{kernelss}
\mathbf{K}_{ss}(t,t')= \mathbf{Q}_{ss}(t')\delta(t-t')+\mathbf{Q}_{sr}(t)\mathbf{K}_{rs}(t,t')
\end{equation}
and $\delta(t)$ is the \textit{delta} distribution.

The same argument allows us to obtain an inverse relation of \eqref{interact rs} 
\begin{align}\label{interact sr}
\pmb{\varphi}(t)&=\int_\mathbb{R} \mathbf{K}_{sr}(t,t')\pmb{\psi}(t')dt' 
\end{align}
simply interchange in \eqref{kernelrs}, \eqref{evolucion varphi} and \eqref{kernelss} the quantities 
$\pmb{\varphi}\longleftrightarrow \pmb{\psi}$, $r\longleftrightarrow s$.

The mathematical construction presented here shows how a process in $\mathbb{S}$ can be described by the interaction of two sub-processes in $\mathbb{S}_t$ and $\mathbb{R}_t$. Specifically this interaction can be viewed in the relation \eqref{interact rs} or its inverse \eqref{interact sr}. 

In other words, the process on $\mathbb{S}_t$ can be seen as the interaction between the processes on  $\mathbb{S}$ and $\mathbb{R}_t=\mathbb{S-S}_t$, this interaction emerges from the elements of infinitesimal transition probabilities present in $\mathbf{K}_{ss}(t,t')$ through \eqref{kernelss} and \eqref{kernelrs}.

%%%%%%%%%%%%%%%%%%%%%%%%%%%%%%%%%%%%%%%%%%%%%%%%%%%%%%%%%%%%%%%%%%%%%%%%%%%%%%%%%%%%%%%%%%%%%%%%%%%%%%%%

\section*{\sc\textbf{A6} Dilation-Dilution transformation}

The \textit{dilution} operation of $\mathbb{S}_t$ into a larger sample space $\mathbb{S}$, can also be understood as an \textit{dilation} represented by the substitution rule $\mathfrak{n}_t\longmapsto N$.  In a general sense, every \textit{dilution-dilation} transformation involves a renormalization of the distribution obtained above. 

To illustrate this point let us consider two distributions $\{p_n:n\in {\scriptstyle \mathbb{S}}\}$ and $\{P_n:n\in {\textstyle \mathbb{S}}\}$  defined over the sample spaces ${\scriptstyle \mathbb{S}}$ and $\mathbb{S}$ respectively, such that ${\scriptstyle \mathbb{S}} \subset\mathbb{S}$ and both are normalized in each samples spaces. By definition we have
\begin{align}\label{term pos}
\sum_{n\in {\textstyle \mathbb{S}}} P_n = \sum_{n\in {\textstyle\mathbb{S}}-{\scriptstyle \mathbb{S}}} P_n
+\sum_{ n\in {\scriptstyle \mathbb{S}}} P_n \: =1.
\end{align}
Since all terms of \eqref{term pos} are positive, there exist a subset of ${\scriptstyle \mathbb{S}}$ in which the distribution  $P_n$, restricted to ${\scriptstyle \mathbb{S}}$, is smaller than $\{p_n:n\in {\scriptstyle \mathbb{S}}\}$. We can see it in this from
\begin{equation}
\sum_{n\in {\scriptstyle \mathbb{S}}} P_n <\sum_{n\in {\scriptstyle \mathbb{S}}} p_n
\end{equation}

We want to describe the process $\{p_n:n\in {\scriptstyle \mathbb{S}}\}$ through the process of  $\{P_n:n\in {\textstyle \mathbb{S}}\}$ \textit{projecting} the distribution $P_n$ over ${\scriptstyle \mathbb{S}}$. Since there are values of ${\scriptstyle \mathbb{S}}$ for which $P_n$ is less than $p_n$, that projection should amplify $P_n$ to improve the stated description. This amplification corresponds to a dilation transformation.

We had mentioned that the renormalization is performed through a linear time-dependent transformation. As we conserve the linearity of the equation \eqref{evoluciont} and since the original sample space $\mathbb{S}_t$ is time-dependent, the norm must be corrected locally. This \textit{local} transformation is structured as a \textit{gauge group}, specifically a \textit{group of local dilations}.

On the other hand, this transformation can be viewed through a \textit{projection} of the distribution $P_n(t)$ on $\mathbb{S}_t$, which filter a subset of states and renormalizes the distribution. We used the concept of conditional probability to relate and motivate the renormalization through a gauge transformation. To be more clear, we distinguish the gauge transformed distribution $\mathfrak{p}_n(t)$ and the projected distribution $q_n(t)$. 

We used a dilation on a generic sample space $\mathbb{S}=[0,N]$, for a natural number $N$.

Lets consider the following example. If $q_n(t)$ is defined as
\begin{equation}
q_n(t)=\mathbb{P}\big[X(t)=n|X(t)\in \mathbb{S}_t\big]
\end{equation}
this distribution is equal to zero for all state such that $n\in \mathbb{S-S}_t$. Applying the Bayes identity \cite{Feller} we have
\begin{equation}\label{valles}
\mathbb{P}[A|B]\mathbb{P}[B]=\mathbb{P}[B|A]\mathbb{P}[A]
\end{equation}
we can express $q_n(t)$ as
\begin{equation}\label{opdef}
q_n(t)=\Lambda(t) P_n(t),
\end{equation}
where \begin{equation}\label{lambda}
\Lambda(t)=\frac{\mathbb{P}\big[X(t)\in \mathbb{S}_t|X(t)=n]}{\mathbb{P}\big[X(t)\in \mathbb{S}_t\big]}.
\end{equation}
The numerator of \eqref{lambda} is only equal to $0$ or $1$ depending on whether $n\in \mathbb{S-S}_t$ or $n\in \mathbb{S}_t$, respectively. The denominator of \eqref{lambda} is $\mathbb{P}\big[X(t)\in \mathbb{S}_t\big]=\mathbb{P}\big[X(t)\leq \mathfrak{n}_t\big]$ and for the mutually exclusive events we have 

\begin{equation}
\mathbb{P}\big[X(t)\in \mathbb{S}_t\big]=\sum_{n\in \mathbb{S}_t} P_n(t)<1
\end{equation}
this implies  $q_n(t)\geq P_n(t)$, for $n\in\mathbb{S}_t$. The expression \eqref{opdef} involves a probability fraction which can be viewed as a projection operator which transforms distributions defined on $\mathbb{S}$ into distributions defined on $\mathbb{S}_t$, by filtering selected states and redefining the correct normalization.

The last discussion about the conditional probability as a projection operation motivates the following construction. Given a process in $\mathbb{S}$ described by $\pmb{\phi}(t)=(P_0(t),\cdots,P_N(t))^\intercal$ which satisfies a general equation, similar to \eqref{evoluciont} 
\begin{equation}\label{ec}
d_t\pmb{\phi}(t)=\mathbf{Q}(t)\pmb{\phi}(t)
\end{equation}
and the dilation transformation
\begin{equation}\label{gauged}
\mathfrak{g}_t:P_{n}(t)\longrightarrow \mathfrak{p}_{n}(t),
\end{equation}
where $\mathfrak{p}_{n}(t)=\lambda(t)P_{n}(t)$ are the components of $\pmb{\phi}'(t)=(\mathfrak{p}_0(t),\cdots,\mathfrak{p}_N(t))^\intercal$. We showed that the covariance requirement is satisfied if $\mathfrak{g}_t$ is a \textit{gauge transformation}, this implies the addition of a covariant derivative. To prove this statement we transform a general evolution equation \eqref{ec} through a local dilation group $\mathfrak{g}_t$
\begin{align}
%d_t\pmb{\phi}'(t)&=d_t\lambda(t)\pmb{\phi}(t)+\lambda(t)d_t\pmb{\phi}(t)\\
d_t\pmb{\phi}'(t) - d_t\lambda(t)[\lambda(t)]^{-1}\pmb{\phi}'(t)&=\mathbf{Q}(t)\pmb{\phi}'(t).
\end{align}
In order to keep the shape of \eqref{evoluciont} when we transformed by $\mathfrak{g}_t$, we need to introduce an affine connection which can be implemented through a covariant derivative given by 
\begin{equation}\label{cov der}
D_t=d_t-\omega(t)
\end{equation}
where $\omega(t)=d_t\lambda(t)[\lambda(t)]^{-1}$ is the \textit{gauge function}.

In this case, the evolution equation for $\pmb{\phi}'(t)$ is 
\begin{align}\label{evolucion gauged}
D_t\pmb{\phi}'(t)=\mathbf{Q}(t)\pmb{\phi}'(t)
\end{align}
which is invariant under the local dilation group.

The definition \eqref{cov der} allows the distribution $P_n(t)$ and $\mathfrak{p}_n(t)$, related by gauged corresponds to a Markov process. 

For the case of this work, the gauge function $\omega(t)$ is independent of the $t-$generation, then $\lambda(t)=2^{\mathfrak{a}t+\mathfrak{b}}$, for $\mathfrak{a}, \mathfrak{b}$ are real constants.
Otherwise, the time dependence of $\omega(t)$ introduces a time inhomogeneity in the process.

The example used to motivate the definition \eqref{gauged} can be extended to describe a rich variety of cases. If we generalized $q_n(t)$ as 
\begin{equation}\label{primer general}
q_n(t)=\mathbb{P}\big[X(t)=n|X(t)\in \mathbb{S}_t,Z\big],
\end{equation}
where $Z$ is an extra condition and eventually space-time dependent. The distribution \eqref{primer general} is equal to zero for $n\in \mathbb{S-S}_t$. Then for the Bayes identity \eqref{valles} applied to  \eqref{primer general}
\begin{equation}\label{opdef2}
q_n(t)=\Lambda_n(t) P_n(t)
\end{equation}
where \begin{equation}
\Lambda_n(t)=\frac{\mathbb{P}\big[ X(t)\in \mathbb{S}_t ,Z|X(t)=n]}{\mathbb{P}\big[X(t)\in \mathbb{S}_t,Z\big] }.
\end{equation}
\vspace*{0.2cm}

In a more abstract sense,  we consider a partition of $\mathbb{S}=\{B_n: n\in \mathbb{S}\}$ and let us consider the event $A_n$ write in the following way: 
\begin{equation}
A_n=\bigcup_{m\in\mathbb{S}} A_n\cap B_m.
\end{equation}

Therefore we use the law of total probability  
\begin{equation}\mathbb{P}[A_n]=\sum_{m\in\mathbb{S}}\mathbb{P}[A_n|B_m]\mathbb{P}[B_m]\end{equation}for a particular case of the partition $B_m=\{m\}$ and $\mathbb{P}[A_n]=q_n(t)$, then \begin{equation}\label{ultima ecuacion}q_n(t)=\sum_{m\in\mathbb{S}}\Lambda_{nm}(t) P_m(t)\end{equation}
where 
\begin{equation}
\Lambda_{nm}(t)=\mathbb{P}[A_n|X(t)=m].
\end{equation}

But of course, the expression \eqref{ultima ecuacion} contains the example presented in this work if we take $\Lambda_{nm}(t)\equiv\lambda(t)\delta_{nm}$. The equation \eqref{ultima ecuacion} motivates the natural generalization of a gauge transformation proposed in \eqref{Gauge GENERAL}.

%Correspondence and request for materials should be addressed to M. Caruso mcaruso@ugr.es and C. Jarne jarne@fisica.unlp.edu.ar.
%\bibliography{aps-ancestrosNotes}

\end{document}